# On the relationship between classes P and NP

Anatoly Plotnikov

Department of "Computer Systems and Networks"
of the Dalh East-Ukrainian National University,
Luhansk, 91034, Ukraine.
E-mail: a.plotnikov@list.ru

**Abstract**

In this paper we discusses the relationship between the known classes P and NP. We show that the difficulties in solving problem "P versus NP' have methodological in nature. An algorithm for solving any problem is sensitive to even small changes in its formulation. As we will shown in the paper, these difficulties are exactly in the formulation of some problems of the class NP.

## 1 Introduction

Unsuccessful attempts to develop an efficient (polynomial-time) algorithm for many practically important problems raised the question of the formalization of a class of problems that have be solved. This formalization led to the concept of the class NP — the class problems solvable by a nondeterministic Turing machine [3].

Eliminating from consideration the problems that obviously can not be solved in polynomial time, is an important step in determining the class of problems for which it makes sense to look for an effective algorithm. It is such problems as finding a set of all subsets of a finite set, construction of all independent sets of a graph, etc.

For practical purposes it is convenient to use the following definition of the class NP [5]:

A problem $Z$ belongs to the class NP if:

1. the problem can be defined by a finite number $n$ of symbols;

2. the problem solution can be represented by a finite number $m$ of symbols, where $m$ is a polynomial function of $n$: $m = f(n)$;

3. the time $t$ for verifying the obtained solution is some polynomial functions of $n$: $t = \varphi(n)$.

Each problem of NP is solvable in the classical sense, since it can be solved by a deterministic Turing machine [3, 2].

Solution algorithm for problem $Z \in$ NP is called *effective* if solution of $Z$ can be obtained in the polynomial number of steps of $n$. The set of problems of NP, having the polynomial-time solution algorithm, forms a class of P, where P $\subseteq$ NP.

Unfortunately, there are the problems of NP, for which the effective solution algorithm is unknown. Analysis of problems of the class NP shows that the main difficulty is the need to perform exhaustive search of solution elements.

The following questions arise: What causes are the exhaustive search? When it makes sense to seek an effective solution algorithm?

These questions raise the problem of the relationship between classes P and NP, that is, proving one of the relations: P= NP or P $\neq$ NP. Attempt to solve this problem is the subject of many works. A brief history of these attempts to solve this problem can be found at "P– versus – NP" page of G. J. Woeginger (see [10]).

In this paper we show that the difficulties in solving problem of the relationship between classes P and NP have methodological in nature. As it is well known [9], an algorithm for solving any problem is sensitive to even small changes in its formulation. As it will shown below, these difficulties are exactly in the formulation of problems of the class NP.

To solve this problem, it is enough to formulate a question: whether the problem $Z \in$ NP is solvable *in this setting*? We specify the properties problems of NP, that require a complete inspection of the elements of the solution and we determine the class of problems of NP, for which it is reasonable to seek an effective solution algorithm. Then the other problem of NP are defined as the exponential in nature. Consequently, for such problems, finding an effective solution algorithm does not make sense in the *current statement*.

## 2 Mathematical model

Let $R$ is a finite set and $Q = \{R_1, R_2, \ldots, R_m\}$ is a family of nonempty subsets of $R$. A pair $(R, Q)$, satisfying the property

$$\text{if } R_1 \in Q \text{ and } R_2 \subset R_1 \text{ then } R_2 \in Q \qquad (1)$$

called a *hereditary system*, or an *independent system*.

Let a problem $Z$, belonging to the class NP, is given on a hereditary system. Then, each of the subsets of $R_j \in Q$ $(j = 1, 2, \ldots, m)$ is called *an admissible solution* of the problem $Z$, and every maximal respect to inclusion $R_j$ is called *support solution* of the problem. In the problem $Z$ it is required to find a support solution that satisfies the given conditions.

The most important question that arises in the process of solving the problem, consists in determining the membership of any subset of $R_k \subset R$ to the set of admissible solutions of $Q$.

Consider some examples of such problems.

**Satisfiability problem (SAT).** Let $\pi$ is a Boolean function over $n$ variables $x_1, \ldots, x_n$, represented in conjunctive normal form, i.e. $\pi$ is a conjunction of clauses. It is required to find a satisfying assignment of variables, i.e. such that $\pi$ is "true".

Let $R$ is the set of literals $r$, where $r$ is either $x_i$ or $\bar{x}_i$ $(i = \overline{1, n})$, belonging to some clause of the $\pi$. Then the set of admissible solutions $Q$ contains such a subset of set $R$, each of which has literals of some clauses of $\pi$. To solve the question "$R_j \in Q$?" $(R_j \subset R)$, it needs to determine that:

- the subset of $R_j$ contains no alternative literal, that is, a variable $x_i$ or $\bar{x}_i$ simultaneously;

- there are clauses of $\pi$, containing together all literals from a subset $R_j$.

It is clear that the SAT problem is defined on a hereditary system.

**The Hamiltonian cycle problem.** Let $G = (V, E)$ is $n$-vertex undirected graph. It is required to find a cycle of graph edges of $G$, which includes each of $n$ vertices exactly once.

We show that this problem is also defined on a hereditary system.

Let $S$ is the set of Hamiltonian cycles $R^*$ of $G$. Denote a set $Q$ of subsets of edges of $R_j$ $(j = 1, 2, \ldots, m)$ of $E$ such that $R_j \in Q$, if and only if $R_j \subseteq R^*$, $(R^* \in S)$. Obviously, to answer on the question "$R_j \in Q$?", it needs to determine that:

- a subset of $R_j$ does not induce the subgraph whose vertices have degrees more than two;

- there exists a Hamiltonian cycle of $R^* \in S$ such that $R_j \subseteq R^*$.

Clearly, the two $(E, Q)$ is a hereditary system.

Let $(R, Q)$ is a hereditary system.

Let, further, $w(r_i)$ $(i = \overline{1, n})$ is an integer, called the *weight* of element $r_i \in R$. For each $R_j \in Q$ we define the sum

$$w(R_j) = \sum_{\forall r \in R_j} w(r).$$

This sum is called a *weight* of the subset $R_j$.

Suppose we want to find $R^* \in Q$, which has the maximum weight. In this case, we have formulated the optimization problem.

**The maximum independent set problem (MMIS).** Let $G = (V, E)$ is a $n$-vertex undirected graph. It is required to find a subset $V^* \subseteq V$, which has the maximum number of vertices such that each pair of vertices in $V^*$ is not adjacent in $G$.

We establish that this problem is also defined on a hereditary system.

Any subset of $R_j \subseteq V$ is called *independent* if every two vertices of $R_j$ are non-adjacent. Let $Q$ is the set of all independent sets of $G$. It is easy to see that the pair $(V, Q)$ is a hereditary system. In this problem, we have the weight $w(R_j) = Card(R_j)$ for any $R_j \in Q$. To solve the problem "$R_j \in Q$?" ($R_j \subset V$) it is sufficient to show that vertices of $R_j$ are pairwise non-adjacent.

It follows from the above examples that the problem of the class NP should be considered as a four $(R, Q, \mathcal{P}, f)$, where $R$ is the set of all elements of the solution, $Q$ is the set of admissible solutions of the problem, $\mathcal{P}$ is the system of predicates which determine that a $R_j \subset R$ is a admissible solution, i.e. $R_j \in Q$ and $f$ is the map: $R_j \to W$ that defines the "weight" of the admissible solution. In the future, we believe that $f(R_j)$ can be computed in polynomial time from value $Card(R_j)$.

Thus, we have defined a *set-theoretic model* of a problem of NP. In computational complexity theory, each problem of NP is considered as a problem *recognition*. A recognition problem is a computational problem, whose solution is "*yes*" or "*no*" [4]. The solution of computational problem (in this case, a support solution) can be considered as "proof" that the corresponding recognition problem of NP has an answer "yes". Therefore, the concept of an admissible solution is broader than the concept of "proof" for the recognition problem.

## 3 Sequential method

Let there be a problem $Z \in$ NP. We assume that $Z = (R, Q, \mathcal{P}, f)$ is determined on a hereditary system $(R, Q)$. The question arise: *How* can we construct an admissible solution $Z$?

A Turing machine is a general model of computation (see, for example, [1, 4]). Therefore, we can assume that we have a single-tape Turing machine $M$. The machine $M$ processes symbols in the cells of the tape *sequentially*, i.e. symbol by symbol. If we assume that the Turing machine solves the problem $Z$ then we can consider the result of each step of $M$ as an admissible solution $Z$. It is natural to consider recording a symbol on the tape as the constructing the next element of the admissible solution.

Thus, the procedure constructing any admissible solution $R_j \in Q$ ($j = 1, 2, \ldots, m$) is extensive in the time, i.e. its elements are obtained sequentially, element by element.

The method of constructing the required solution, when we obtain its elements step by step, element by element, we call *sequential* [6, 8].

Let $R_1$, $R_2 \in Q$ is admissible solutions of the problem $Z \in$ NP such that $R_1 \subset R_2$. We denote the time of their constructing as $t(R_1)$ and $t(R_2)$ respectively. Then the following assertion takes place.

**Theorem 1** $t(R_1) < t(R_2)$.

**Proof.** Without loss of generality, we can assume that $Card(R_2) = Card(R_1)$ +1. By definition of the sequential method, the admissible solution $R_2$ can be obtained *after* the construction by a Turing machine of the admissible solution $R_1$. This will require at least one cycle of the machine. This proves Theorem 1. Q.E.D.

**Theorem 2** *The solution of any problem $Z \in$ NP can be obtained by a sequential method.*

**Proof.** We believe that every problem of the class NP is solvable, i.e. each such problem can be solved by a deterministic Turing machine. Since this machine runs sequentially, it produces the solution step by step, an element of element. Therefore, Theorem 2 is true. Q.E.D.

Obviously, we can assume that the sequential method is solely the *general* method for solving each problem $Z \in$ NP.

Indeed, for example, suppose we need to find some independent set graph vertices. Any independent set of graph vertices is an admissible solution of MMIS. However, only the maximal independent set is the support solution of MMIS. The global solution of this problem — the maximum independent set — is also some support solution.

Obviously, in common case, the *simultaneous choice* of several independent vertices is not possible if the structure of the graph is not known *in advance*. It is clear that every next subsequence of vertices can be selected only if we know what vertices have already been selected in an independent set earlier.

## 4 Problems without foresight

Let there be a problem $Z \in$ NP, determined on the hereditary system $(R, Q)$. Let, further, $R_1 \in Q$ is an admissible solution that is not support. In accordance with forming the next admissible solution $R_2$ such that $R_1 \subset R_2$, and $Card(R_2) = Card(R_1) + 1$, the problem of NP can be divided into two classes [7, 6]:

- the problems for which the next solution $R_2 = R_1 \cup \{r\}$ ($r \in R$) can be found in the polynomial time by means of joining to the $R_1$ of one element of the set $R \setminus R_1$;

- all the other problems of NP.

In other words, the problem of NP can be classified according to the computation time of the predicate "$R_1 \cup \{r\} \in Q$?" for any admissible solution $R_1 \in Q$ and for each element $r \in R \setminus R_1$. If such predicate can be computed in the polynomial time from the dimension of the problem then this problem is called a *problem without foresight*. Otherwise, the problem is called *exponential in nature*.

Note that the selected term — problem that is exponential in nature — should not be taken literally. As is usual in the complexity theory (see, for example, [3]) such problems may require $O(n!)$ or $O(n^n)$ steps to compute the predicate "$R_1 \cup \{r\} \in Q$?".

The set of all problems without foresight will be denoted by UF, where UF $\subseteq$ NP.

**Theorem 3** *The support solution $Z \in$ NP can be found in the polynomial time if and only if $Z \in$ UF.*

**Proof.** Let there be a problem $Z \in$ NP such that $Z \in$ UF. By the definition of problems without foresight, each next admissible solution of $Z$ can be found in the polynomial time. Since $\oslash \in Q$ for any problem $Z \in$ UF and the support solution contains no more than $p(n)$ of elements, where $n$ is the dimension of the problem, and $p(n)$ is a polynomial, this implies the polynomial time of building the support solution.

On the other hand, let there is the problem $Z \in$ NP is such which is solvable in the polynomial time. Assume that $Z \notin$ UF. In this case, there exists at least one admissible solution of $Z$, found in the exponential time. By condition of Theorem 3, the support solution of problem $Z \in$ UF is found in the polynomial time. We have a contradiction of Theorems 1 and 2. Q.E.D.

Thus, the class UF induced problems of NP for which the support solution can be constructed in the polynomial time. In common case, each support solution is not the desired solution of the problem.

**Theorem 4** *UF $\subset$ NP and UF $\neq$ NP.*

**Proof.** By definition of UF $\subseteq$ NP. In the Hamiltonian cycle problem, it needs to determine what the constructed set of edges belongs to at least one Hamiltonian cycle, which requires to foresee of this fact and therefore requires exhaustive search of elements of the solution, that is, UF $\neq$ NP. Q.E.D.

Taking into account Theorems 3 and 4, we have the following result.

**Corollary 1** $P \subset UF$ and $P \neq NP$.

# 5 Conclusion

Using our terminology, we can reformulate the third item in definition of the class NP as follows:

3. time $t$ for verifying the obtained *support solution* is some polynomial function from $n$: $t = \varphi(n)$.

Then we have following definition of the class UF. A problem $Z$ belongs to a class UF If:

1. the problem can be defined by a finite number $n$ of symbols;

2. the problem solution can be represented by a finite number $m$ of symbols, where $m$ is a polynomial function of $n$: $m = f(n)$;

3. the time $t$ for verifying *any admissible solution* is some polynomial functions of $n$: $t = \varphi(n)$.

As we can see the definition of the class UF differs from the definition of the class NP only the third item. The definition of the class NP wider than the problems of the class UF since the class NP can includes and objectives for which the time for verifuing any admissible solutions is exponentially, i.e. requires a lot of exhaustive search. Thus, the class NP includes all problems of the class UF. It is easy to see that the class of P is entirely in the class UF.

Thus, a direct answer to the problem of "P vs NP" consists in the fact that the class P is not equal to the class NP.

When the question of the relationship between classes P and NP was formulated, many researchers believed that this will solve the problem on the possibility to develop (or not) an effective (polynomial-time) algorithm for all problems of the class NP. However, we found that the existing definition of class NP is redundant, i.e. it allows to include in this class of problems that are exponential in nature. As we earlier removed the problems with exponential length solutions from the class NP, now, we remove from the class of NP the problems that are exponential in nature.

In other words, it is expedient to focus on the study of problems without foresight (the class UF) for searching an effective solution algorithm.

However, in this case, we have two questions.

The first question concerns the problems which are exponential in nature. We see that such practical important problems as the Hamiltonian cycle problem (in the current formulation) can not be solved effectively. However, it is necessary take into account that the set of problems without the foresight includes such NP-complete problems as the satisfiability problem and the maximum independent set problem. However, every problem of the class NP can be reduced to some (NP-complete) problem without foresign in the polynomial time, for example, to the satisfiability problem [3]. Such reduction can be considered as reformulation of the relevant problem that is exponential in nature into the problem without the foresight.

In particular, to find a Hamiltonian cycle in a graph, one can formulate the following optimization problem: find a partition of the graph into the minimum number of disjoint cycles [8].

The second question consists that we must be reformulated the old problem of "P vs NP" into a new one, namely: "P vs UF".

Thus, unfortunately, the elucidation of the relationship between classes P and NP does not answer on the question about possibility of effectively solving problems as the class NP, and the class UF. We just found that the lack of an existing definition of the class NP consists that we do not determine the polynomial time of checking the truth of the predicate $\mathcal{P}$ for the problem $(R, Q, \mathcal{P}, f) \in$ NP.

To have a hope for constructing effective solution algorithm, we need consider the problem of the class UF. Obviously, that the determining properties of these problems is a possibility to construct a support solution in the polynomial time. If the problems $Z_1$ is out the class UFthen we must reduce it into a $Z_2$ from UF, the global solution of which is also a global solution $Z_1$.

Sometimes we can not say that a problem belongs to UF or not. In this case, we have little knowledge about this problem to determine its existence in the class UF. Then it is expedient to suspect that the problem lies outside of the class UF.